\documentclass[%
reprint,
superscriptaddress,
 amsmath,amssymb,
 aps,
pre,
twocolumn
]{revtex4-2}

\usepackage{graphicx}
\usepackage{dcolumn}
\usepackage{bm}

\usepackage{placeins}
\usepackage{xcolor}

\usepackage[normalem]{ulem}

\newcommand{\abs}[1]{\left\lvert#1\right\rvert}
\newcommand{\diff}{d} 
\newcommand{\e}{\textup{e}}
\newcommand{\avg}[1]{\left\langle#1\right\rangle}

\newcommand{\figabrev}{Fig.}
\newcommand{\multfigabrev}{Figs.}
\newcommand{\eqabrev}{Eq.}
\newcommand{\secabrev}{Sec.}
\newcommand{\refabrev}{Ref.}
\newcommand{\multirefabrev}{Refs.}

\renewcommand{\vec}[1]{\mathbf{#1}}
\usepackage{nicefrac}
\usepackage{microtype}
\usepackage{hyperref}
\usepackage{tikz}
\usetikzlibrary{math}
\usetikzlibrary{external}
\begin{document}

\title{Superadiabatic Dynamical Density Functional Theory for One-Dimensional Brownian Hard-Rod Fluids}

\author{Jens Weimar}
\email{jens.weimar@uni-tuebingen.de}
\affiliation{Institute for Applied Physics, University of Tübingen, Tübingen, Germany}
\author{Daniel de las Heras}
\affiliation{Institute for Theoretical Physics, University of Tübingen, Tübingen, Germany}
\author{Martin Oettel}
\affiliation{Institute for Applied Physics, University of Tübingen, Tübingen, Germany}

\date{\today}

\begin{abstract}
We study the relaxation dynamics of a one-dimensional system of hard rods at the level of one-body fields using superadiabatic dynamical density functional theory, and compare the results with Brownian dynamics simulations. The theory uses the exact equilibrium excess free energy functional and incorporates superadiabatic effects from first principles via an adiabatic closure relation at the three-particle level. The predicted superadiabatic response shows very good agreement with simulation data, provided that initial-state ensemble differences between theory and simulations remain small. Furthermore, we evaluate an approximate power functional theory based on a velocity-gradient expansion, which also yields an accurate superadiabatic response, although requiring a memory kernel extracted from simulation data.
\end{abstract}

\maketitle


\section{Introduction}
Classical density functional theory (DFT) provides a formally exact route to the
equilibrium microstructure and thermodynamics of classical many-particle systems
\cite{bobstart}. Its dynamical extension, dynamical density functional theory
(DDFT) \cite{ddft1,ddft2,ddft3} in its standard formulation, does not inherit this status.
DDFT rests on the adiabatic approximation, in which the instantaneous
nonequilibrium state is treated as an equilibrium state carrying the same
one-body density, so that the dynamics is a sequence of equilibrium states and
memory effects are absent \cite{ddftfrompof}. Contributions beyond this approximation --- termed
superadiabatic --- have been shown to be substantial and, in many situations,
dominant \cite{superadiabaticeffects}.

Superadiabatic contributions are accounted for exactly by power functional
theory (PFT) \cite{pftstart,pftmegascript,pftcorrection,pftentropy}, which is to the dynamics what DFT is to equilibrium. Unlike the equilibrium case, however, no exact power functional is known, and the theory must be closed by an approximation. Two routes have been pursued: analytical ansätze \cite{sinuspftfit,pftmemory,pftmixtures} for the excess power functional like one based on gradients of the one-body velocity field, enhanced with a multiplicative memory kernel \cite{sinuspftfit}, and functionals learned by machine learning from steady-state simulation data \cite{pftml1,pftml2}.
The analytical approximations demonstrate strong performance within their valid regimes. However, they incorporate undetermined transport coefficients (e.g., shear and bulk viscosities) that must be inferred from experimental or simulation results, or derived via alternative theoretical methods.
Machine learning functionals trained on steady-state data can be used to model fully non-equilibrium dynamics~\cite{pftml2}, though memory effects are then only captured approximately.
A scheme that generates superadiabatic forces from first principles is therefore desirable.

Such a scheme was proposed by Tschopp and Brader
\cite{sddft1,sddft2,sddft3}. Starting from force DDFT
\cite{forcedft}, in which the adiabatic two-body density is constructed
explicitly from the inhomogeneous Ornstein-Zernike equation rather than from a
one-body correlation function, the adiabatic closure is deferred to the
three-body level.
The resulting superadiabatic DDFT (SDDFT) couples the
evolution of the one- and two-body densities and produces an approximated superadiabatic current that, notably,
contains no free parameters and requires no spatial or temporal assumptions.
SDDFT has so far been applied to
three-dimensional hard spheres in planar geometry \cite{sddft1,sddft2}
and to a steady state of two-dimensional hard disks \cite{sddft3}, with
promising results. In none of these cases is the exact equilibrium functional
known, so that residual deviations from simulation cannot be attributed
unambiguously to the three-body closure rather than to the underlying functional
approximation.

This ambiguity motivates the present study. 
For one-dimensional hard rods, the exact grand canonical functional is known~\cite{percusfunctional}.
In principle, this eliminates errors arising from the equilibrium functional, leaving the closure as the sole approximation of the scheme.
Superadiabatic effects in this system are strong
\cite{superadiabaticeffects}, and the relaxation of rods released from a periodic external
potential has already been analysed within PFT \cite{sinuspftfit}, providing
a direct point of comparison. One-dimensional systems are, however, also the
least favourable setting with respect to a second and independent source of
error: differences between the canonical ensemble realised in simulation and the
grand canonical ensemble underlying the functional
\cite{canonicaldftfromdecomposition,canonicalddftfromdecomposition,canonicalddft2,orderpreservingwittmann}. These differences decay with
increasing particle number but can be pronounced for the strongly confined,
few-particle, configurations typical of one-dimensional test cases. We will
see that ensemble differences set the accuracy limit of the comparison rather than the closure
itself.

We find that SDDFT reproduces the Brownian dynamics (BD) density evolution
almost quantitatively in a setup chosen such that ensemble differences are
small, in contrast to DDFT, which relaxes too fast. The
superadiabatic current, which opposes the relaxation and exceeds the adiabatic
current in magnitude, is likewise captured, with deviations smaller than those
obtained using the simplest analytical PFT~\cite{sinuspftfit}.
The remaining deviations between the superadiabatic currents in SDDFT and BD are of comparable size and
opposite sign to those of the adiabatic currents, which identifies them as an
artefact of the ensemble mismatch rather than a shortcoming of the closure. When
ensemble differences are large, as for expansion from a parabolic trap, all theoretical
schemes considered here fail to reproduce the simulation results.

The paper is organised as follows. Since neither force DDFT nor SDDFT is an
established standard, \secabrev~\ref{sec:theory} recapitulates their derivation from the
Smoluchowski equation. \secabrev~\ref{sec:1d} specialises the equations to
one-dimensional hard rods, describes their numerical solution, and summarises
the PFT approximation \cite{sinuspftfit} and the BD simulations used for
comparison. \secabrev~\ref{sec:results} presents the density evolution and the splitting of the
current into adiabatic and superadiabatic contributions. \secabrev~\ref{sec:conclusions} concludes and gives an outlook.

\section{Theory}
\label{sec:theory}

We start with a self-contained summary for the derivation of the basic (S)DDFT equations. For more details we refer to  Refs.~\cite{forcedft,sddft1}.
We start with the Smoluchowski equation for the time evolution of the $N$-particle probability function $P(\vec{r}^N,t)$ (where $\vec{r}^N=\{\vec{r}_1,\dots,\vec{r}_N\}$ and $t$ denotes time) in a Brownian system:
\begin{equation} \label{eq:smoluchowski}
    \begin{aligned}
        \frac{1}{D_0} \, \frac{\partial P(\vec{r}^N,t)}{\partial t}
        = \sum_{i=1}^{N} \vec{\nabla}_{\vec{r}_i} \cdot \bigl( & P(\vec{r}^N,t) \, (\vec{\nabla}_{\vec{r}_i} \ln(P(\vec{r}^N,t)) \\
        & + \vec{\nabla}_{\vec{r}_i} \, \beta \, U(\vec{r}^N,t)) \bigr),
    \end{aligned}
\end{equation}
Here, $D_0$ is the bare diffusion coefficient, $\beta = 1/(k_B T)$ is the inverse temperature,
$\nabla_{\vec{r}_i}$ is the derivative with respect to the position of particle $i$, and, assuming only two-body interactions, the configurational energy $U(\vec{r}^N,t)$ is given by
\begin{equation}
    U(\vec{r}^N,t) = \sum_{i=1}^{N} \sum_{j=i+1}^{N} \phi(r_{ij}) + \sum_{i=1}^{N} V^\text{ext} (\vec{r}_i,t),
\end{equation}
with $V^\text{ext}$ the (in general time-dependent) one-body external potential and $\phi(r_{ij})$ the pair potential between particles with $r_{ij} = \abs{\vec{r}_i - \vec{r}_j}$ being the inter-particle distance.
Although nonconservative external forces can be incorporated into SDDFT \cite{sddft3}, we neglect them here for simplicity.
Integrating \eqabrev~\eqref{eq:smoluchowski} over $N-1$ particle coordinates yields an equation for the averaged one-body density profile $\rho(\vec{r}_1,t)$:
\begin{equation}  \label{eq:oneparticledynamics}
    \begin{aligned}
        \frac{1}{D_0} \, \frac{\partial \rho (\vec{r}_1,t)}{\partial t} = \vec{\nabla}_{\vec{r}_1} \! \cdot \! \Bigl(& \vec{\nabla}_{\vec{r}_1} \, \rho (\vec{r}_1,t) \\
        & + \rho (\vec{r}_1,t) \, \vec{\nabla}_{\vec{r}_1} \, \beta \, V^\text{ext} (\vec{r}_1,t) \\
        & + \int \rho_2 (\vec{r}_1,\vec{r}_2,t) \, \vec{\nabla}_{\vec{r}_1} \, \beta \, \phi (r_{12}) \, \diff \vec{r}_2 \Bigr).
    \end{aligned}
\end{equation}
DDFT rests upon the so-called adiabatic approximation: the time-dependent two-particle density $\rho_2 (\vec{r}_1,\vec{r}_2,t)$ is taken as the equilibrium two-particle density  $\rho_2^\text{ad}(\vec{r}_1,\vec{r}_2,[\rho])$ of an equilibrium system having a one-body density, $\rho^\text{eq}$, given by the instantaneous density of the non-equilibrium system, $\rho^\text{eq}=\rho(\vec{r}_1,t)$. Therefore, $\rho_2^\text{ad}$ is a functional of the instantaneous density. The condition of a fixed density profile in the adiabatic system defines a corresponding adiabatic one-body external potential which fulfills:
\begin{equation}
c_1 (\vec{r}_1) = \ln (\rho (\vec{r}_1)) + \beta \, V^\text{ad} (\vec{r}_1) - \beta \, \mu\,,\label{eq:c1}
\end{equation}
and, after taking the gradient,
\begin{equation} \label{eq:adiabaticpotential}
    - \vec{\nabla}_{\vec{r}_1} (\beta \, V^\text{ad} (\vec{r}_1,[\rho])) = \vec{\nabla}_{\vec{r}_1} \ln(\rho(\vec{r}_1,t)) - \vec{\nabla}_{\vec{r}_1} c_1(\vec{r}_1,[\rho])\,.
\end{equation}
Here $c_1(\vec{r}_1,[\rho])$ is the first-order direct correlation function.
Equation~\eqref{eq:c1} is the Euler-Lagrange equation of the equilibrium density functional and Eq.~\eqref{eq:adiabaticpotential} is the equilibrium one-body force balance equation of the system.

\subsection{``Standard'' (or potential) DDFT}

This formulation uses the functional definition of $c_1$ from the excess free energy functional, $\mathcal{F}^\text{ex}[\rho]$:
\begin{equation} 
  c_{1,p}(\vec{r}_1,[\rho])=- \beta \, \frac{\delta \mathcal{F}^\text{ex}[\rho]}{\delta \rho(\vec{r}_1)}.\label{eq:c1potential}
\end{equation}
Here, as in \refabrev~\cite{forcedft}, the index $p$ refers to potential DDFT. Furthermore, it uses the equilibrium sum rule \cite{bobstart}
\begin{equation} \label{eq:c1pot}
    \vec{\nabla}_{\vec{r}_1} \, c_{1,p}(\vec{r}_1,[\rho]) = - \int \frac{\rho_2^\text{ad}(\vec{r}_1,\vec{r}_2,[\rho])}{\rho(\vec{r}_1)} \, \vec{\nabla} (\beta\phi (r_{12})) \, \diff\vec{r}_2 \;.
\end{equation}
Inserting \eqabrev~\eqref{eq:c1pot} into the adiabatic approximation for Eq.~(\ref{eq:oneparticledynamics}) yields the standard or potential DDFT equation
\begin{equation} \label{eq:onepaticledynamicsc1}
    \begin{aligned}
        \frac{1}{D_0} \, \frac{\partial \rho (\vec{r}_1,t)}{\partial t} = \vec{\nabla}_{\vec{r}_1} \! \cdot \! \Bigl(& \vec{\nabla}_{\vec{r}_1} \, \rho (\vec{r}_1,t) \\
        & + \rho (\vec{r}_1,t) \, \vec{\nabla}_{\vec{r}_1} \, \beta \, V^\text{ext} (\vec{r}_1,t) \\
        & - \rho (\vec{r}_1,t) \vec{\nabla}_{\vec{r}_1} \, c_{1,p} (\vec{r}_1,[\rho]) \Bigr)\;.
    \end{aligned}
\end{equation}
In practice, this equation is used with an approximation of the excess free energy functional $\mathcal{F}^\text{ex}[\rho]$ for the system of interest, which is equivalent to a direct approximation for the one-body function $c_1$. 

\subsection{Force DDFT}

Here, the starting point is the explicit adiabatic two-particle density which is written as 
\begin{equation} \label{eq:twoparticledensityconstrunction}
    \rho_2^{\text{ad},f} (\vec{r}_1,\vec{r}_2,[\rho]) = \rho(\vec{r}_1) \, \rho(\vec{r}_2) \, (1+ h(\vec{r}_1,\vec{r}_2,[\rho])),
\end{equation}
where we have added an index $f$ to make clear that $\rho_2^{\text{ad},f}$ is computed in the force DDFT scheme via explicit, inhomogeneous 2-point correlations. 
Equation~\eqref{eq:twoparticledensityconstrunction} uses the total correlation function, $h(\vec{r}_1,\vec{r}_2,[\rho])$, which is computed via the inhomogeneous Ornstein-Zernike (OZ) equation 
\begin{equation}
    \begin{aligned}
        h(\vec{r}_1,\vec{r}_2,[\rho]) = & \, c_2(\vec{r}_1,\vec{r}_2,[\rho]) \\
        & + \int h(\vec{r}_1,\vec{r}_3,[\rho]) \, \rho(\vec{r}_3) \, c_2(\vec{r}_3,\vec{r}_2,[\rho]) \, \diff \vec{r}_3.
    \end{aligned}
    \label{eq:oz}
\end{equation}
Here, $c_2 (\vec{r}_1,\vec{r}_2,[\rho])$ is the second-order direct correlation function, which is defined as 
\begin{equation}
  c_2 (\vec{r}_1,\vec{r}_2,[\rho]) = -\beta \, \frac{\delta^2 \mathcal{F}^\text{ex}[\rho]}{\delta \rho(\vec{r}_1)\delta \rho(\vec{r}_2)}.\label{eq:c2force}
\end{equation}
Note that both $h$ and $c_2$ are functionals of the inhomogeneous one-body density $\rho$.
In practice, an approximate functional for calculating this \textit{two}-body quantity is used, or an integral equation closure linking $h$ and $c_2$ is employed to solve Eq.~(\ref{eq:oz}) \cite{forcedft3}. Having obtained $h$ and thus $\rho_2^\text{ad}$, the equilibrium sum rule (\ref{eq:c1pot}) is used to calculate a force-based $c_1$ explicitly:  
\begin{equation} \label{eq:c1force}
    \vec{\nabla}_{\vec{r}_1} \, c_{1,f}(\vec{r}_1,[\rho]) = - \int \frac{\rho_2^{\text{ad},f}(\vec{r}_1,\vec{r}_2,[\rho])}{\rho(\vec{r}_1)} \, \vec{\nabla} (\beta\phi (r_{12})) \, \diff\vec{r}_2,
\end{equation}
with again the index $f$ referring to force DDFT.  As before, inserting \eqabrev~\eqref{eq:c1force} into the adiabatic approximation for \eqabrev~(\ref{eq:oneparticledynamics}) yields the force DDFT equation
\begin{equation} \label{eq:oneparticledynamicsc1f}
    \begin{aligned}
        \frac{1}{D_0} \, \frac{\partial \rho (\vec{r}_1,t)}{\partial t} = \vec{\nabla}_{\vec{r}_1} \! \cdot \! \Bigl(& \vec{\nabla}_{\vec{r}_1} \, \rho (\vec{r}_1,t) \\
        & + \rho (\vec{r}_1,t) \, \vec{\nabla}_{\vec{r}_1} \, \beta \, V^\text{ext} (\vec{r}_1,t) \\
        & - \rho (\vec{r}_1,t) \vec{\nabla}_{\vec{r}_1} \, c_{1,f} (\vec{r}_1,[\rho]) \Bigr),
    \end{aligned}
\end{equation}
which differs from the potential DDFT equation (\eqabrev~\eqref{eq:onepaticledynamicsc1}) only by the use of $c_{1,f}$ instead of $c_{1,p}$. If the exact $\mathcal{F}^\text{ex}$ is used, then $c_{1,f}=c_{1,p}$ and force and potential DDFT are identical. If an approximate functional is used, then it is used in either the approximation of the \textit{two}-body direct correlation function in force DFT, \eqabrev~\eqref{eq:c2force}, or the approximation of the \textit{one}-body direct correlation function in potential DFT, \eqabrev~\eqref{eq:c1potential}, which leads to different results \cite{forcedft,forcedft4,forcedft2}.

\subsection{Superadiabatic DDFT (SDDFT)}

To include superadiabatic effects in force DDFT, one needs to go beyond the adiabatic approximation for the time-dependent two-particle density. This can be done by making the adiabatic approximation at the next higher order, i.e., the three-particle level \cite{sddft1,sddft2}.
To this end, we reconsider the Smoluchowski equation, \eqabrev~\eqref{eq:smoluchowski}, and integrate over $N-2$ particle coordinates, which yields
\begin{equation} \label{eq:smoluchowskin-2}
    \begin{aligned}
        \frac{1}{D_0} \, & \frac{\partial \rho_2 (\vec{r}_1,\vec{r}_2,t)}{\partial t} \\
        = \sum_{i=1,2} \vec{\nabla}_{\vec{r}_i} \cdot \Bigl( & \vec{\nabla}_{\vec{r}_i} \, \rho_2 (\vec{r}_1,\vec{r}_2,t) \\
        & + \rho_2(\vec{r}_1,\vec{r}_2,t) \, \vec{\nabla}_{\vec{r}_i} \beta \, (V^\text{ext}(\vec{r}_i,t) + \phi(r_{12})) \\
        & + \int \rho_3 (\vec{r}_1,\vec{r}_2,\vec{r}_3,t) \, \vec{\nabla}_{\vec{r}_i} \, \beta \, \phi(r_{i3}) \, \diff \vec{r}_3 \Bigr).
    \end{aligned}
\end{equation}
Here, $\rho_3$ is the three-particle density.
In equilibrium, $\rho_2 \to \rho_2^\text{ad}$, $\rho_3 \to \rho_3^\text{ad}$ (both depend functionally on $\rho$) and $V^\text{ext} \to V^\text{ad}$ (which is the adiabatic external potential generating the one-body density $\rho$, given in \eqabrev~\eqref{eq:adiabaticpotential}) and the time derivative on the left hand side of \eqabrev~\eqref{eq:smoluchowskin-2} vanishes. We now subtract the equilibrium version of this equation from the dynamic one. In the dynamic equation, the adiabatic approximation on the three-particle level is used:  
\begin{equation}
    \rho_3 (\vec{r}_1,\vec{r}_2,\vec{r}_3,t) \approx \rho_3^\text{ad} (\vec{r}_1,\vec{r}_2,\vec{r}_3,[\rho])
\end{equation}
such that the last term of \eqabrev~\eqref{eq:smoluchowskin-2} is the same in the equilibrium and the dynamical versions, and therefore cancels in the difference. This yields the equation for the evolution of the two-particle density \cite{sddft1}:
\begin{equation} \label{eq:twoparticledensityevolution}
    \begin{aligned}
        \frac{1}{D_0} & \frac{\partial \rho_2(\vec{r}_1,\vec{r}_2,t)}{\partial t} \\ 
        = \sum_{i=1,2} \vec{\nabla}_{\vec{r}_i} \cdot \Big( & \vec{\nabla}_{\vec{r}_i} \,   \rho_2(\vec{r}_1,\vec{r}_2,t) - \vec{\nabla}_{\vec{r}_i} \,   \rho_2^\text{ad}(\vec{r}_1,\vec{r}_2,[\rho])\\
        & + \rho_2(\vec{r}_1,\vec{r}_2,t) \, \vec{\nabla}_{\vec{r}_i} \beta \phi(r_{12}) \\
        & - \rho_2^\text{ad}(\vec{r}_1,\vec{r}_2,[\rho]) \, \vec{\nabla}_{\vec{r}_i} \beta \phi(r_{12}) \\
        & + \rho_2(\vec{r}_1,\vec{r}_2,t) \, \vec{\nabla}_{\vec{r}_i} \beta V^\text{ext}(\vec{r}_i,t) \\
        & - \rho_2^\text{ad}(\vec{r}_1,\vec{r}_2,[\rho]) \, \vec{\nabla}_{\vec{r}_i} \beta V^\text{ad}(\vec{r}_i,[\rho]) \Big)
    \end{aligned}    
\end{equation}
The SDDFT scheme is complete by solving this equation for $\rho_2(\vec{r}_1,\vec{r}_2,t)$ together with the general evolution equation for $\rho(\vec{r},t)$ (\eqabrev~\eqref{eq:oneparticledynamics}). Thus, it is a coupled system of partial integro-differential equations, and it is this coupling which is expected to describe memory effects beyond the usual adiabatic approximation of DDFT \cite{ddftklatt}.

\section{Relaxation in 1D}
\label{sec:1d}

So far, SDDFT has been implemented for 3D hard spheres in planar geometry \cite{sddft1,sddft2} and for a steady state of 2D hard disks \cite{sddft3}, with promising results. In both cases, the exact equilibrium density functional (needed to generate $\rho_2^{\text{ad},f}(\vec{r}_1,\vec{r}_2,[\rho])$ via $c_2(\vec{r}_1,\vec{r}_2,[\rho])$) is not known, and therefore it is difficult to ascertain whether the residual differences between theory and simulation have their origin in the adiabatic approximation for the three-particle density  or simply in the approximation for the functional. This is the motivation for considering one-dimensional systems of hard rods (of length $\sigma$) where the exact grand canonical density functional is known \cite{percusfunctional}.
Specifically, we investigate the relaxation of hard rods initially in equilibrium and confined by some external potential which is  switched off at time zero. Such a setting has already been investigated using power functional theory~\cite{sinuspftfit}, aiming at the elucidation of superadiabatic effects. This gives us additionaly the opportunity to compare SDDFT and PFT. In Secs.~\ref{sec:sddft1d} and \ref{sec:pft} we give the necessary theoretical expressions for SDDFT and PFT, respectively, before we discuss results in \secabrev~\ref{sec:results}.

\subsection{Superadiabatic DDFT for 1D Hard Rods} \label{sec:sddft1d}

For the actual implementation of (superadiabatic) force DDFT for 1D hard rods, consider first the evolution equation for $\rho(\vec{r},t)$ (\eqabrev~\eqref{eq:oneparticledynamics}), in particular the last term, which contains a the spatial derivative of the hard pair potential.
Using an approach similar to \cite{forcedft} for hard spheres in planar geometry, this term can be written as
\begin{equation} \label{eq:integralexpression} 
    \begin{aligned}
        & \int_{-\infty}^{\infty} \rho_2 (x_1,x_2,t) \, \frac{\diff}{\diff x_1} (\beta \, \phi_{12}) \, \diff x_2 \\
        =  & \, \rho_2 (x_1,x_1 + \sigma,t) - \rho_2 (x_1,x_1 - \sigma,t).
    \end{aligned}
\end{equation}
For the derivation see Appendix \ref{sec:derivedc1force}.
Using \eqabrev~\eqref{eq:integralexpression}, the evolution of the one-particle density becomes
\begin{equation} \label{eq:oneparticledynamics1d}  
    \begin{aligned}
        \frac{1}{D_0} \, \frac{\partial \rho (x_1,t)}{\partial t} = \frac{\partial}{\partial x_1} \biggl(& \frac{\partial \rho (x_1,t)}{\partial x_1} + \rho (x_1,t) \, \beta \, \frac{\partial V^\text{ext} (x_1,t)}{\partial x_1} +  \\
        & \rho_2 (x_1,x_1 + \sigma,t) - \rho_2 (x_1,x_1 - \sigma,t)
        \biggr).
    \end{aligned}
\end{equation}
If the adiabatic two-particle density is used ($\rho_2 \to \rho_2^{\text{ad},f})$, \eqabrev~\eqref{eq:oneparticledynamics1d} is also the dynamic equation for force DDFT for 1D hard rods.

The evolution equation for the two-particle density in one dimension becomes
\begin{equation} \label{eq:twoparticledensityevolution1d}
    \begin{aligned}
        \frac{1}{D_0} & \frac{\partial \rho_2(x_1,x_2,t)}{\partial t} \\ 
        = \sum_{i=1,2} \frac{\partial}{\partial x_i} \biggl( & \frac{\partial \rho_2^\text{sup}(x_1,x_2,t,[\rho])}{\partial x_i} \\
        & + \rho_2^\text{sup}(x_1,x_2,t,[\rho]) \, \beta \frac{\diff \phi(\abs{x_1 - x_2})}{\diff x_i} \\
        & + \rho_2(x_1,x_2,t) \, \beta \frac{\partial V^\text{ext}(x_i,t)}{\partial x_i} \\
        & - \rho_2^{\text{ad},f} (x_1,x_2,[\rho]) \, \beta \, \frac{\diff V^\text{ad}(x_i,[\rho])}{\diff x_i} \biggr).
    \end{aligned}    
\end{equation}
Here we introduced the superadiabatic two-particle density as
\begin{equation}
    \rho_2^\text{sup} (x_1,x_2,t,[\rho]) = \rho_2 (x_1,x_2,t) - \rho_2^{\text{ad},f} (x_1,x_2,[\rho]).
\end{equation}

As described in \cite{sddft1}, for hard interactions between the particles, the term in \eqabrev~\eqref{eq:twoparticledensityevolution1d} which includes the derivative of the pair potential $\phi$ should be replaced by a zero-flux condition. This means that instead of calculating the time derivative of the two-particle density directly, first, a two-particle flux is calculated without the term involving the pair potential:
\begin{equation}
    \begin{aligned}
        j_i(x_1,x_2,t) = & - D_0 \, \frac{\partial \rho_2^\text{sup}(x_1,x_2,t,[\rho])}{\partial x_i} \\
        & - D_0 \, \rho_2(x_1,x_2,t) \, \beta \frac{\partial V^\text{ext}(x_i,t)}{\partial x_i} \\
        & + D_0 \, \rho_2^{\text{ad},f} (x_1,x_2,[\rho]) \, \beta \, \frac{\diff V^\text{ad}(x_i,[\rho])}{\diff x_i}
    \end{aligned}
\end{equation}
Second, the fluxes directly into and out of the forbidden area of $\rho_2$ (i.e.,\ when the particles would overlap) are set to zero. Finally, the time derivative of the two-particle density is calculated via the continuity equation
\begin{equation}
    \frac{\partial \rho_2(x_1,x_2,t)}{\partial t} = - \sum_{i=1,2} \frac{\partial j_i (x_1,x_2,t)}{\partial x_i}.
\end{equation}
In 3D planar geometry, this is complicated because the grids needed for the solution of the inhomogeneous OZ equation and those optimal for enforcing the zero-flux condition do not match \cite{sddft1}. For pure 1D systems, we do not have this problem. Instead, we can even make a further optimization. In order to enforce particle conservation, it is generally advisable to use a staggered grid for densities and fluxes, where the fluxes are calculated at the interfaces of the density bins, as is common in finite volume methods in fluid dynamics \cite{leveque}. This also simplifies the employment of the zero-flux condition as there are now defined flux grid cells at the interface of the forbidden $\rho_2$ area which can be set to zero without additional side effects on fluxes near the interface.

The adiabatic potential appears in \eqabrev~\eqref{eq:twoparticledensityevolution1d} through its derivative, which using \eqabrev~\eqref{eq:adiabaticpotential} is given by:
\begin{equation} \label{eq:adiabaticpotential1D}
    - \beta \, \frac{\diff V^\text{ad} (x,[\rho])}{\diff x} =  \frac{\partial \ln \rho(x,t)}{\partial x} - \frac{\diff c_1(x,[\rho])}{\diff x}\,.
\end{equation}
The last term involves $c_1$, and this is actually the only occurrence of $c_1$ in the SDDFT scheme. In its original formulation \cite{sddft1}, the force-based expression $c_{1,f}$ was used, which is consistent as the equation for the one-particle dynamics, \eqabrev~\eqref{eq:oneparticledynamics1d}, is also force-based. Since in our system both $c_{1,f}$ and $c_{1,p}$ (obtained from the derivative of the functional) should be equal, we can also use $c_{1,p}$ as an alternative; we call this a hybrid scheme.
\begin{figure}
    \centering
    \includegraphics[width=1\linewidth]{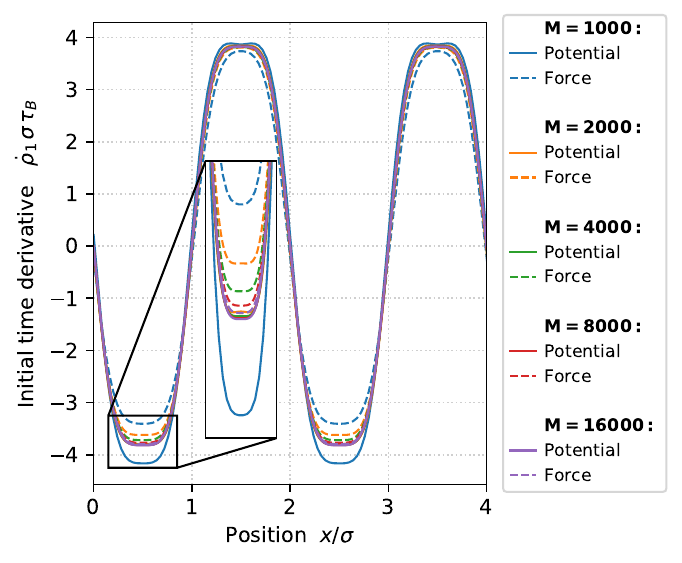}
    \caption{Comparison of potential and force DDFT for the initial density profile $\rho(x)\,\sigma=0.35 \, \sin(40\, \pi \, x/L)+0.45$  in the absence of an external potential. Shown is the initial time derivative of the density $\dot{\rho}_1 \, \sigma \, \tau_B$, with $\tau_B = \sigma^2/D_0$, for different spatial discretizations $\Delta x$, as indicated in the label by the number of bins $M$. The predictions of potential and force DDFT converge as $\Delta x$ decreases. The periodic computational domain is $[0,40\,\sigma)$; for example, $M=4000$ bins correspond to $\Delta x=0.01\,\sigma$. For clarity, only the interval $[0,4\,\sigma]$ is displayed.
    }
    \label{fig:comparisonpotentialforceddft}
\end{figure}

We consider systems of length $L=40$ $\sigma$, and the spatial axis is discretized into $M$ bins. Equations~\eqref{eq:oneparticledynamics1d} and \eqref{eq:twoparticledensityevolution1d} are solved using a Runge-Kutta scheme \cite{rkdormandprince,rkshampine,scipy} with a variable timestep. At each time step, the computation of $\rho_2^{\text{ad},f}$ in the force DDFT and SDDFT schemes necessitates an inversion of an $M \times M$ matrix to solve the inhomogeneous OZ equation, \eqabrev~\eqref{eq:oz}. Thus, small values of $M$ are highly desirable. However, there are discretization effects which we illustrate here through a comparison of potential DDFT and force DDFT which should deliver identical results using the exact Percus functional for 1D hard rods. Figure~\ref{fig:comparisonpotentialforceddft} shows the time derivative $\partial \rho(x,t)/\partial t|_{t=t_0}$ for an arbitrarily chosen instantaneous profile $\rho(x,t_0)$ for potential and force DDFT with the number of bins varying from $M=1000$ to 16000. The difference between the schemes varies much with the specific situation, but the difference between potential and force DDFT consistently decreases as the number of bins increases.

However, the notable differences seen here in the instantaneous derivative are less prominent in the time-integrated profile $\rho(x,t)$. Here, already for $M=4000$ ($\Delta x = 0.01\sigma$) the results for potential and force DDFT are almost identical and this choice is used in the following. A finer discretization is numerically very costly as this not only leads to an increase in the computational cost of solving the inhomogeneous OZ equation but also to much smaller time steps, since $\Delta t \propto (\Delta x)^2$.

For the comparison with PFT below, it is useful to introduce the one-particle current, $j_\text{tot}(x,t)$, and split it into an ideal part, $j_\text{id}$, an external part, $j_\text{ext}$, an adiabatic part, $j_\text{ad}$, and a superadiabatic part, $j_\text{sup}$. That is,
\begin{equation} \label{eq:jsplitting}
    j_\text{tot}(x,t) = j_\text{id}(x,t) + j_\text{ext}(x,t) + j_\text{ad}(x,t) + j_\text{sup}(x,t).
\end{equation}
The ideal current is given by the diffusion of an ideal gas
\begin{equation} \label{eq:jid}
    j_\text{id} (x_1,t) = - D_0 \, \frac{\partial \rho (x_1,t)}{\partial x_1}.
\end{equation}
The external current is given by
\begin{equation} \label{eq:jext}
    j_\text{ext} (x_1,t) = - D_0 \, \rho(x_1,t) \, \beta \, \frac{\partial V^\text{ext} (x_1,t)}{\partial x_1},
\end{equation}
which, however, is always zero in our relaxation scenarios, as we remove the external potentials at $t=0$.
The adiabatic current is given by the free energy contribution. It is the same as in force DDFT
\begin{equation} \label{eq:jad}
    j_\text{ad} (x_1,t) = - D_0 \left(\rho_2^\text{ad} (x_1,x_1 + \sigma,t) - \rho_2^\text{ad} (x_1,x_1 - \sigma,t)\right).
\end{equation}
The superadiabatic current describes contributions that go beyond the adiabatic approximation and are also originated by the interparticle interactions. It is similar to the adiabatic current, but using the superadiabatic two-particle density
\begin{equation} \label{eq:jsup}
    j_\text{sup} (x_1,t) = - D_0 \left(\rho_2^\text{sup} (x_1,x_1 + \sigma,t) - \rho_2^\text{sup} (x_1,x_1 - \sigma,t)\right),
\end{equation}
This splitting is introduced for the comparative analysis below. In the actual numerical scheme, one uses the total two-particle density and does not split it into the adiabatic and superadiabatic contributions.
Note that the definition of all of these currents is the same for both the force and the hybrid versions as they only differ in the equation for the two-particle current.

\subsection{Power Functional Theory}

\label{sec:pft}

We use the power functional approximation of Ref.~\cite{sinuspftfit} based on a gradient expansion of the velocity profile. 
We simply outline here the main features of the framework and refer the reader to Ref.~\cite{sinuspftfit} for futher details.

In power functional theory for Brownian dynamics~\cite{pftstart}, the time evolution of the one-body fields is obtained via miminimization of a functional, $R[\rho,j]$, with units of power, which depends functionally on the density and the current profiles. The dependency is on the history of both fields from the initial time, in which the system is assumed to be in equilibrium, to the current time. At each time, the functional is minimized with respect to the physical current while keeping the density profile fixed.
To formulate analytical approximations, it is convenient in general, to express the functional using the density and the velocity profiles~\cite{sinuspftfit,delasHeras2020}.
Density and current profiles are linked via the continuity equation.
Hence, as shown in Ref.~\cite{pftcorrection} it is also possible to write down the functional using only the current profile.

Minimization of the functional produces its associated Euler-Lagrange equation, which is, by construction, the exact one-body force-density balance equation. In 1D, it reads
\begin{equation}
\gamma j = -k_BT\frac{\partial\rho}{\partial x} +\rho f_{\text{ext}} + \rho f_{\text{int}},\label{eq:jtot}
\end{equation}
where, for clarity, we have omitted the spatio-temporal $(x,t)$ dependency on all fields. Here, $\gamma=1/(\beta D_0)$ is the friction coefficient against the implicit solvent, the first term in the right hand side is the (ideal gas) diffusive transport term, $f_{\text{ext}}(x,t)$ is an external force field, and $f_{\text{int}}(x,t;[\rho,v])$ is the internal force field originated by interparticle interactions. The later inherits the functional dependecy of the generating functional. Here, we use $v=j/\rho$, the velocity profile. It is convenient to split the internal force field into adiabatic (equilibrium-like) and superadiabatic (genuine non-equilibrium) contributions:
\begin{equation}
f_{\text{int}}(x,t;[\rho,v]) = f_{\text{ad}}(x,t;[\rho]) + f_{\text{sup}}(x,t;[\rho,v]).
\end{equation}
The adiabatic contribution depends functionally only on the density profile, and as in DDFT, is given by the equilibrium excess free-energy functional
\begin{equation}
f_{\text{ad}}(x,t;[\rho]) = - \frac{\partial}{\partial x}\left(\frac{\delta \cal{F}^{\text{ex}}[\rho]}{\delta\rho(x,t)}\right).
\end{equation}
Here $\cal{F}^{\text{ex}}$ is evaluated at each time at the true non-equilibrium profile. The superadiabatic internal force field is given by the functional derivative of the excess power functional
\begin{equation}
f_{\text{sup}}(x,t;[\rho,v]) = - \frac{1}{\rho(x,t)}\left.\frac{\delta P^{\text{ex}}[\rho,V]}{\delta v(x,t)}\right|_{V=v},\label{eq:fsup}
\end{equation}
where, $V$ denotes a trial velocity field and $v$ the actual physical velocity field. Again, the density profile is kept fixed while performing the functional derivative. Following Ref.~\cite{sinuspftfit} we approximate $P^{\text{ex}}$ by the leading term of an expansion in the velocity gradient, which for 1D system reduces to
\begin{equation}
P^{\text{ex}}=k_BT \, \frac{k(t)}{2}\int \rho(x,t)\left[\frac{\partial V(x,t)}{\partial x}\right]^2dx.
\end{equation}
Here, $k$ is an unknown memory kernel which in general can also carry a dependence on the overall density.
The superadiabatic force follows from Eq.~\eqref{eq:fsup}
\begin{equation}
f_{\text{sup}}(x,t;[\rho,v]) = \frac{k_B T \, k(t)}{\rho(x,t)}\frac{\partial}{\partial x}\left[\rho(x,t)\frac{\partial v(x,t)}{\partial x}\right].\label{eq:fsupext}
\end{equation}
The adiabatic and superadiabatic currents are simply the contributions of the corresponding internal force fields to the total current. That is,
\begin{align}
j_{\text{ad}} & = \gamma^{-1}\rho f_{\text{ad}},\\
j_{\text{sup}} & = \gamma^{-1}\rho f_{\text{sup}}.\label{eq:jsup_pft}
\end{align}

If the memory kernel, $k(t)$, in Eq.~\eqref{eq:fsupext} is given, then all terms contributing to the total current, Eq.~\eqref{eq:jtot}, are explicitly known and the dynamics of the density profile follows straightforwardly from the continuity equation. Here, we obtain the optimal kernel from simulations. At each time, we fix the density profile to that of the BD simulations and also set a fixed value for the memory kernel.
We then minimize the complete power functional~\cite{sinuspftfit} with respect to the velocity profile. 
We repeat the minimization for different values of the memory kernel in order to find the kernel that minimizes the differences between the superadiabatic force predicted by PFT and that in simulations. 
That is, we essentially fit the time-dependent amplitude of the superadiabatic response by comparing it to BD simulations.
The spatial structure is given completely by Eq.~\eqref{eq:fsup}.

\subsection{Brownian Dynamics Simulation} \label{sec:bd}
We compare our results {from} the four (S)DDFT schemes (potential and force DDFT, force and hybrid SDDFT) as well as from PFT to results obtained with Brownian dynamics (BD) simulations, where the hard potential between rods is approximated by a soft potential $ \propto r^{-20}$ (similar to Ref.~\cite{sinuspftfit}).
Beyond the density profiles, we also sample the currents from BD and perform the splitting into the ideal, adiabatic, and superadiabatic contributions~\cite{superadiabaticeffects}. 
The spatio-temporally resolved one-body fields, such as density and current profiles, are obtained by averaging over $10^8$ to $10^9$ independent simulation runs. Each run initializes with a distinct microstate sampled from the equilibrium distribution and uses a different noise realization. The high number of independent simulations is both made possible and necessary due to the low number of particles in the system.

\section{Results}
\label{sec:results}

\subsection{Density Evolution}
\begin{figure}[h!]
    \centering
    \includegraphics[width=1\linewidth]{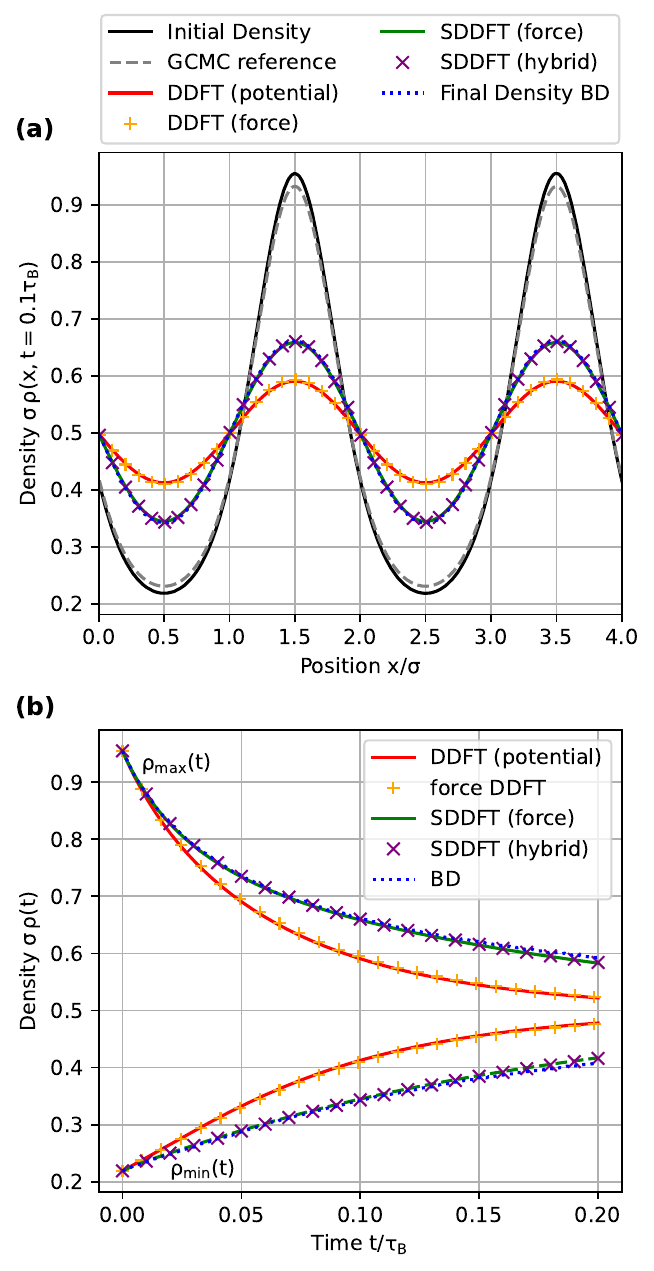}
    \caption{(a) Comparison of BD, potential DDFT, force DDFT, and the force and hybrid variants of SDDFT at $t=0.1\,\tau_B$ during free expansion from the equilibrium BD density profile generated by the external potential in \eqabrev~\eqref{eq:vextsinus}. Only the interval $x\in[0,4\,\sigma]$ of the periodic profile is shown for clarity.
    (b) Time evolution of the minimum and maximum values of the density profile for BD, potential DDFT, force DDFT, and both SDDFT variants. For force DDFT and hybrid SDDFT, only selected time points are shown for clarity. The spacing between the symbols therefore does not represent the numerical time step, which is much shorter.
    }
    \label{fig:densityevolution}
\end{figure}
A sensitive test case for superadiabatic effects is to consider the free expansion of $N$ particles initially {equilibrated} inside a periodic sinusoidal potential with $N$ valleys. {Potentials of this form have already been considered in Ref.~\cite{superadiabaticeffects} (introducing the splitting between adiabatic and superadiabatic contributions) and in Ref.~\cite{sinuspftfit} (using a velocity-gradient expansion for the excess power functional for hard rods) 
which gives us the opportunity to compare our results to another, structurally very different, method to go beyond the adiabatic approximation.
An advantage of using such an $N$-valley, $N$-particle setup is that
differences between the canonical and grand canonical equilibrium densities are small. Our particular choice for the external potential is
\begin{equation} \label{eq:vextsinus}
    \frac{V^\text{ext} (x)}{k_B T}  = V_0^\text{ext} \, \sin \left(\frac{2 \, \pi \, x \, N}{L}\right),
\end{equation}
with $L=40 \,\sigma$, $N = 20$ and $V_0^\text{ext} = 1$. 
In \figabrev~\ref{fig:densityevolution} (a) the initial equilibrium density for fixed $N=20$ (canonical ensemble, from BD) is shown as a solid-black line and the grand-canonical equilibrium density for $\langle N \rangle =20$ (Percus functional) is shown
with a dashed-gray line. Both profiles are rather close. However, the difference is still noticeable and some ensemble differences must be expected in the relaxation process. To minimize these effects further, we initialize all (S)DDFT schemes {(potential and force DDFT, force and hybrid SDDFT)} with the equilibrium BD profile. However, this will not completely eliminate  ensemble differences as the external potential which would create the equilibrium BD density in the grand-canonical ensemble differs from that in \eqabrev~\eqref{eq:vextsinus}, and the underlying excess free energy functional is formulated in the grand canonical ensemble}~\cite{canonicaldftfromdecomposition,canonicalddftfromdecomposition}.

At time $t=0$ the potential is removed and the system starts to relax to the homogeneous state. The density profiles for all four (S)DDFT schemes at time  $t = 0.1 \, \tau_B$ (with $\tau_B = \sigma^2/D_0$) 
are compared to the BD result in \figabrev~\ref{fig:densityevolution} (a).
The curves for both DDFT variants do almost perfectly match, as do the curves for both SDDFT variants.
Furthermore, both DDFT variants show the known overestimation of the relaxation velocity \cite{ddft1,ddft2,ddft3,superadiabaticeffects,sinuspftfit}.
However, most interestingly, SDDFT shows an almost perfect match with the BD curve.
This hints that SDDFT incorporates almost all superadiabatic effects in this 1D system.
To visualize the accuracy of all methods at different times, we show the time evolution of the minimum and maximum values of the density profile in Fig.~\ref{fig:densityevolution} (a).
Here, we again see an almost perfect match between force-based and potential-based methods and we also find a very good match between SDDFT and BD data. Only at the end of the time window, we observe small deviations at the level of a few per cent. These could be residual ensemble effects (as discussed above), accumulated numerical errors caused by the finite discretization or indeed inaccuracies caused by the adiabatic approximation on the three-particle level inherent to SDDFT.
Note that all methods will eventually converge to the same homogeneous equilibrium profile.

These findings persist also for stronger external potentials, such as e.g.~using $V_0^\text{ext} = 3$, which produces higher initial local densities. 
The first row of \figabrev~\ref{fig:currents} shows DDFT and SDDFT profiles at different times after removing the external potential and compares them to BD results. Again, we see an almost perfect match for SDDFT while DDFT is again too fast.

Note that ensemble differences can have a huge influence on the results \cite{canonicaldftfromdecomposition,canonicalddftfromdecomposition,canonicalddft2,orderpreservingwittmann}. We demonstrate this in Appendix \ref{sec:ensembledifferences}, where we apply all (S)DDFT schemes to a scenario with an initial parabolic potential where there is already a big difference between the canonical and grand-canonical initial profiles. In this case, all four schemes fail to deliver the correct density evolution.

\subsection{Adiabatic and Superadiabatic Currents} \label{sec:currents}
\begin{figure*}[t]
    \centering
    \includegraphics[width=1.0\linewidth]{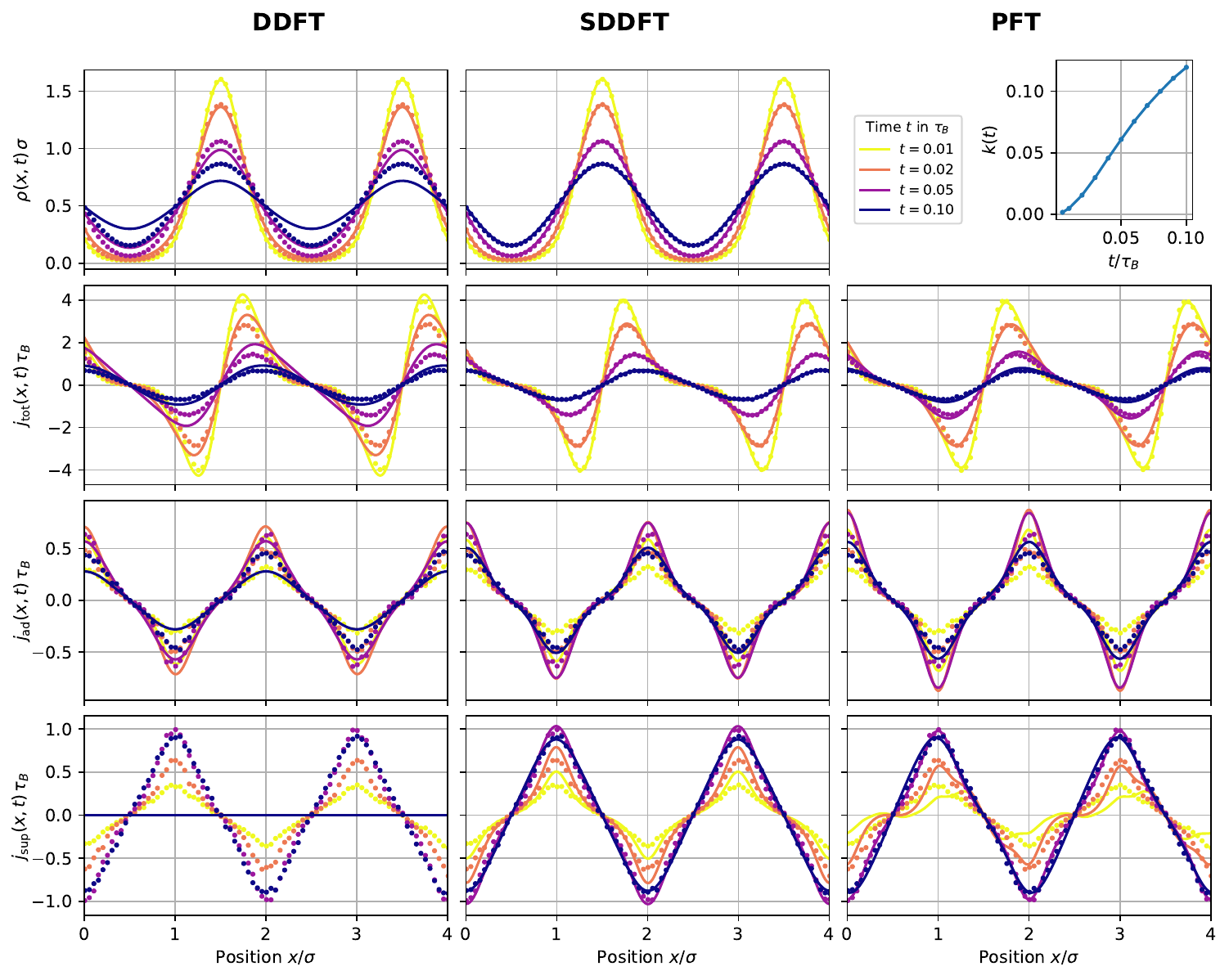}
    \caption{Comparison of potential DDFT, force SDDFT, and PFT for the relaxation of a density profile after removal of the external potential defined in \eqabrev~\eqref{eq:vextsinus} at time $t=0$. The system contains $N=20$ particles in a periodic domain of length $L=40\,\sigma$, with $V_0^\text{ext}=3$. The first row shows the self-consistent density evolution predicted by DDFT and SDDFT (lines), compared with BD data (dots). PFT does not provide a self-consistent density evolution here because $k(t)$ is obtained from a fit to the BD density; the resulting $k(t)$ is shown in the inset. The second row shows the total current, $j_\text{tot}=j_\text{id}+j_\text{ad}+j_\text{sup}$, for DDFT, SDDFT, approximate PFT, and BD at the same times. The DDFT and SDDFT currents are evaluated using their self-consistent densities, whereas the PFT currents are evaluated using the BD densities. The third and fourth rows show the corresponding adiabatic and superadiabatic currents, respectively. The adiabatic PFT current is essentially the potential-DDFT current evaluated using the BD density, while DDFT has no superadiabatic contribution. BD currents are shown as dots throughout. Only the interval $[0,4\,\sigma]$ is displayed. Apart from numerical uncertainties, all profiles are periodic over the full domain of length $40\,\sigma$. Here, SDDFT denotes the force variant; the hybrid variant is visually indistinguishable.
    }
    \label{fig:currents}
\end{figure*}

Analyzing the currents gives further insights.
Aside from the density evolution in the first row, \figabrev~\ref{fig:currents} shows the total ($j_\text{tot}$, second row), adiabatic ($j_\text{ad}$, third row), and superadiabatic ($j_\text{sup}$, {fourth} row) currents for DDFT {(first column)}, SDDFT {(second column)} and  PFT {(third column)} and compares them to BD simulations {(dots in every plot)}.

As one would expect from the almost perfect match of the density evolution of SDDFT to the BD data, there is also an almost perfect match for $j_\text{tot}$ obtained from SDDFT and BD simulations. In contrast, noticeable differences are visible for the total current in BD and DDFT since the later ignores all superadiabatic effects.
At first glance, the agreement between DDFT and BD appears to improve over time.
However, this is merely an artifact of error cancellation between the ideal and adiabatic currents.
Note that the DDFT and BD density profiles diverge significantly by the latest time depicted.
The total current $j_\text{tot}$ from the approximated PFT is also close to that in simulations
but SDDFT recovers the BD results with higher accuracy. 

The current can be further analyzed through its ideal, 
adiabatic and superadiabatic contributions, see \eqabrev~\eqref{eq:jsplitting} (here $j_\text{ext}=0$).
From these, the ideal contribution, $j_{\text{id}}=-D_0\rho'$, is tied to the evolving density in a simple way and does not reveal new insights about the different schemes and will not be discussed further.  

The adiabatic contribution $j_\text{ad}$ (third row of \figabrev~\ref{fig:currents}) is a small contribution to $j_\text{tot}$ and promotes relaxation of the profile towards an homogeneous system.
In practice, however, results do differ somewhat even if the underlying densities are almost identical. This can be seen in the difference for $j_\text{ad}$ from PFT and SDDFT (with nearly equal $\rho(x,t)$) and reflects the difference between potential DDFT (used for $j_\text{ad}$ in PFT) and force DDFT  (used for $j_\text{ad}$ in force and hybrid SDDFT).
More interestingly, differences in $j_\text{ad}$ are visible when comparing either PFT or SDDFT to BD data. These discrepancies are due to residual ensemble differences.
For the BD data, the adiabatic-superadiabatic split is done inside the canonical ensemble, while all other schemes use the grand canonical ensemble.
For instance, in the case of SDDFT by splitting the two-particle density into an adiabatic and superadiabatic part via the grand canonical inhomogeneous OZ equation.
Ensemble differences, and hence discrepancies in the predicted $j_{\text{ad}}$, are larger at short times and vanish as the system approaches equilibrium.

The last row in \figabrev~\ref{fig:currents} shows $j_\text{sup}$, the superadiabatic current which is zero for DDFT.
The supeardiabatic current opposes the adiabatic current, delaying therefore the relaxation of the density profile.
Its magnitude is considerably larger than that of $j_\text{ad}$. Both PFT and SDDFT do capture $j_\text{sup}$ accurately. However, as pointed out in \secabrev~\ref{sec:pft} and \secabrev~\ref{sec:sddft1d}, the methodology to obtain $j_{\text{sup}}$ differs substantially in both schemes.
PFT employs an ansatz for the spatiotemporal dependence of $j_\text{sup}$ in a trial functional which is minimized with respect to the velocity profile.
The functional contains an unknown position independent memory kernel, $k(t)$, which effectively sets the magnitude of the superadiabatic response, and that needs to be fitted from simulation data.
We show the kernel in the upper-right panel of Fig.~\ref{fig:currents}.
The simple PFT approximation used here reproduces the simulation results rather accurately.
For short times, the overall PFT prediction still shows deviations, but it improves rapidly at longer times.
In contrast, for SDDFT the emergence of a non-vanishing superadiabatic response is a direct consequence of the adiabatic closure at the three-particle level, and does not involve unknown kernels that need to be adjusted with simulation data.
For short times, the comparison with BD shows deviations, but they are considerably smaller than in the case of PFT.
For longer times, SDDFT and BD essentially agree (as does PFT). 

SDDFT is is an approximated theory, so one could expect some deviations for the superadiabatic current. However, we again have to consider ensemble differences, as the superadiabatic current is calculated from the superadiabatic two-particle densities with \eqabrev~\eqref{eq:jsup}. The splitting between superadiabatic and adiabatic two-particle densities is done using the grand canonical inhomogeneous OZ equation. The deviations  between the superadiabatic current in SDDFT and BD are roughly of the same magnitude as those for the adiabatic current and are largely in the opposite direction. Therefore, it is plausible that the bulk of the deviations is not intrinsic to the scheme, but just a consequence of residual ensemble differences.

\section{Conclusion and Outlook}
\label{sec:conclusions}

\textit{Summary and conclusion: -- }
Superadiabatic DDFT is a closed theory for the time evolution of the one-body density which needs the free energy functional of the system and employs an adiabatic approximation for three-particle correlations. We have tested it in systems of hard rods in 1D to describe relaxation after an initially confining external potential has been turned off. In 1D, the exact grand canonical density functional of hard rods is known, which eliminates potential sources of inaccuracy for this part of the theory. However, as the dynamic evolution proceeds in the canonical ensemble, a residual inaccuracy remains, related to the difference between canonical and grand canonical functionals. For test cases where these differences are presumably small, we obtain an excellent description of the relaxation dynamics which captures the large superadiabatic effects very well. Small differences between simulation and theory at longer relaxation times are presumably attributable to the residual ensemble effect. For initially confining potentials which show a large difference between the starting canonical and grand canonical density profiles, the ensemble differences dominate and neither standard DDFT nor SDDFT captures the relaxation correctly.   

We have compared our results to PFT \cite{sinuspftfit}, which formulates the time evolution as a minimization problem at every instant in time. The employed ansatz for the excess power functional contains one term generating the adiabatic current and another term generating the superadiabatic current.
The later is approximated by the leading term in an expansion in the 1D velocity gradient, multiplied by a time-dependent memory kernel.
The power functional is minimized with respect to the one-particle velocity field while keeping the exact time-dependent density profile obtained from simulations fixed. The memory kernel is constructed from the simulation data.
The superadiabatic current shows good agreement with simulations, performing only slightly worse than SDDFT.
A direct comparison between the theoretical expressions for the superadiabatic currents in SDDFT and PFT is not possible. 
In SDDFT, the superadiabatic current is written in terms of the superadiabatic two-particle density, Eq.~\eqref{eq:jsup}, whereas in PFT it is written as a functional of the one-body density and velocity profiles, Eq.~\eqref{eq:jsup_pft}.

\textit{Outlook: -- }
Our results show that SDDFT is an excellent candidate for a successful dynamic theory but in the current 1D system, ensemble differences are somehow problematic. In principle, however, these could be overcome by employing the decomposition method of \refabrev~\cite{canonicalddftfromdecomposition} which reconstructs canonical observables from grand canonical ones. This would come at the price of a drastic slow down of the solution for the two-particle density, \eqabrev~\eqref{eq:twoparticledensityevolution1d}. There, a canonical adiabatic two-particle density can be constructed from a set of grand canonical adiabatic two-particle densities (each of these necessitates a solution of the inhomogeneous OZ equation which is the bottleneck of the computation). The size of the set is $N$, the number of particles, and thus the method of \refabrev~\cite{canonicalddftfromdecomposition} would increase the computation time by a factor of order $N$, making it longer than the simulation time.

An elegant solution would be to employ a canonical density functional. Subsequently, the inhomogeneous OZ equation could be replaced by its canonical counterpart \cite{canonicaloz}
without any further modifications. However, despite ongoing research \cite{canonicalfunctionallutsko,canonicaldftwhite,canonicaldfthernando,canonicaldftdibernardobrader,canonicalfunctionalml}, no applicable functional is known.

Although the free energy functional is unknown for most interparticle potentials, machine-learned functionals~\cite{sammuellerc1,buiunifiedmachinelearningframework,reviewalessandromartin} have recently emerged as a powerful tool to find very accurate approximations.
Recently, Ram et al.\ showed that such a functional for the 3D Lennard-Jones system can also be successfully applied to DDFT \cite{mlddft}. SDDFT places higher demands on the quality of the learned functional as it also requires a precise $c_2$ to solve the inhomogeneous OZ equation, but with recent advances \cite {sammuellerc1,dftmlpaircorrmatching} this appears feasible.

Superadiabatic Density Functional Theory and Power Functional Theory offer complementary frameworks for describing non-equilibrium many-body dynamics at the one-body level.
The primary strength of SDDFT lies in its self-contained structure since it relies solely on the equilibrium excess free energy functional without requiring additional phenomenological assumptions or free parameters.
In the present application, SDDFT provides excellent accuracy, reliably reproducing the system's superadiabatic response.
However, its main limitations are computational and structural.
Solving the inhomogeneous Ornstein–Zernike equation can become numerically demanding, particularly in two- and three-dimensional flows.
Furthermore, because SDDFT is inherently an approximate scheme based on a three-body adiabatic closure, systematically incorporating higher-order superadiabatic contributions might not be trivial.
Testing SDDFT in complex flow geometries, such as two-dimensional setups where superadiabatic forces separate into distinct viscous and structural contributions~\cite{delasHeras2020}, presents a promising direction for future work. 
PFT is an exact framework capable, in principle, of describing superadiabatic effects under arbitrary non-equilibrium conditions.
For Brownian dynamics, PFT establishes that the internal force field is a unique functional of the spatio-temporal density and velocity profiles.
The main challenge in PFT is that the power functional must be explicitly approximated.
Analytical approximations based on velocity-gradient expansions typically depend on unknown transport coefficients as input.
Alternatively, recently developed neural functionals~\cite{pftml2} learn the functional mapping with high fidelity, though this relies on pre-generating training data via particle-based simulations.
Ultimately, both approaches represent powerful, synergistic routes toward a complete description of driven soft-matter systems.

\section*{Acknowledgments}
We thank Salomée Tschopp for useful discussions. This work is supported by the Deutsche Forschungsgemeinschaft (DFG, German Research Foundation), project 535083866. 
D.dlH. acknowledges support through the Heisenberg program of the DFG under project 550390029.

\appendix

\section{Derivation of \eqabrev~\eqref{eq:integralexpression}} \label{sec:derivedc1force}
To derive \eqabrev~\eqref{eq:integralexpression}, we first consider the pair potential $\phi_{12} = \phi 
(\abs{x_1 - x_2})$. It is infinite if the distance between the particles is smaller than the particles' length, $\sigma$, and zero if it is larger. Thus, we get
\begin{equation}
    \e^{-\beta \, \phi_{12}} = \theta (\abs{x_1 - x_2} - \sigma).
\end{equation}
Similar to \cite{forcedft}, we now use the chain rule
\begin{equation}
        \frac{\diff}{\diff x_1}  \e^{-\beta \, \phi_{12}} 
        = - \e^{-\beta \phi_{12}}  \,  \frac{\diff}{\diff x_1} \left(\beta \, \phi_{12}\right),
\end{equation}
and rearrange it to obtain
\begin{equation}
    \begin{aligned}
        \frac{\diff}{\diff x_1} \left(\beta \, \phi_{12}\right) & = - \e^{\beta \, \phi_{12}} \,  \frac{\diff}{\diff x_1} \left(\e^{-\beta \, \phi_{12}}\right) \\
        & = - \e^{\beta \phi_{12}} \frac{\diff}{\diff x_1} \theta(\abs{x_1 - x_2} - \sigma) \\
        & = \begin{cases}
            -\delta (\abs{x_1 - x_2} - \sigma), & \quad \text{if } x_1 > x_2 \\
            \,\,\,\,\delta (\abs{x_1 -x_2} -\sigma), & \quad \text{if } x_2 > x_1.
        \end{cases}
    \end{aligned}
\end{equation}
Inserting this into the integral expression yields
\begin{equation}
    \begin{aligned}
        & - \int_{-\infty}^{\infty} \rho_2 (x_1,x_2) \, \frac{\diff}{\diff x_1} (\beta \, \phi_{12}) \, \diff x_2 \\
        = & \, \rho_2 (x_1,x_1 - \sigma) - \rho_2 (x_1,x_1 + \sigma).
    \end{aligned}
\end{equation}

\section{System with large ensemble differences} \label{sec:ensembledifferences}

\begin{figure}[h]
    \centering
    \includegraphics[width=\linewidth]{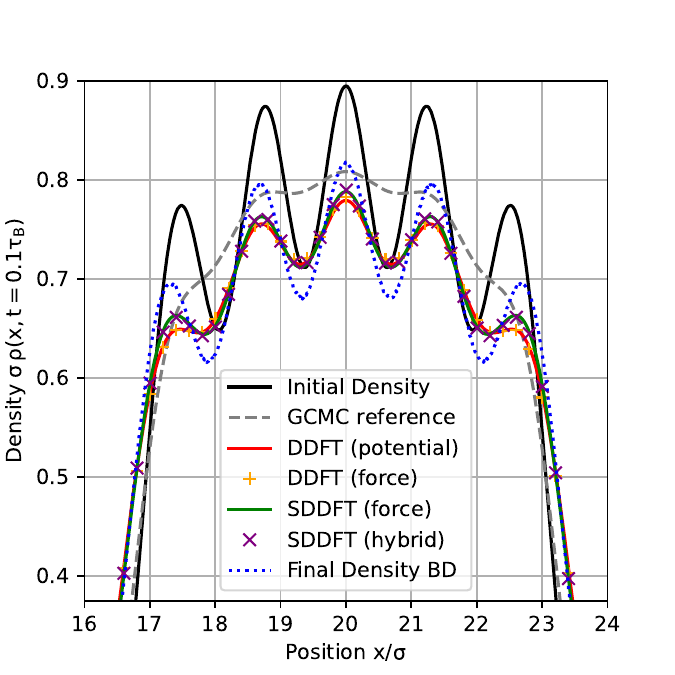}
    \caption{Relaxation of $N=5$ hard rods following the removal of the parabolic external potential $V^\text{ext}(x)=\frac{k_B T}{2\sigma^2}(x-L/2)^2$ at $t=0$. Shown are the canonical and grand-canonical equilibrium density profiles at $t=0$, together with the density profiles at $t=0.1\,\tau_B$ obtained from BD, potential and force DDFT, and the force and hybrid variants of SDDFT.
    }
    \label{fig:catastrophicprofile}
\end{figure}

Ensemble effects can indeed be the main source of the discrepancies still visible in \multfigabrev~\ref{fig:densityevolution} and \ref{fig:currents}. Here we provide an example where ensemble differences affect the results significantly. For that we consider the free expansion of $N=5$ particles from a parabolic confinement.
\figabrev~\ref{fig:catastrophicprofile} shows the initial density obtained with BD simulations (solid-black line) and the grand canonical reference profile (dashed-gray line) with the same average particle number $\avg{N} = 5$.
The difference between the two profiles is very noticeable: the canonical profile shows a distinct layering with five peaks whereas the grand-canonical profile is a fairly unstructured broad peak centered in the middle of the parabolic trap.
The grand canonical density profile is a superposition of canonical profiles with different particle numbers, with the dominant contributions from $N=4$, $5$ and $6$.
In this example, the canonical profile for $N=5$ is completely out of phase relative to those with $N=4$ and $6$. That is, the positions of its peaks and valleys are interchanged.
Thus, the superposition of these profiles
leaves the peaks and valleys of the canonical $N=5$ profile barely visible. The very different form of the density profile also implies that a very different external potential is needed in the grand canonical ensemble to generate the canonical density profile. Consequently, the adiabatic potential from the grand canonical functional (which is used in SDDFT explicitly) is also not correct. 
To show the effect of this, we also plot the density profiles after removing the potential for $t = 0.1 \, \tau_B$, just as in \figabrev~\ref{fig:densityevolution}. Again, we can observe a good match between both DDFT schemes as well as a good match between both SDDFT schemes. Furthermore, SDDFT is again somewhat slower than DDFT. However, all curves are considerably off compared to BD.
Similar problems have been discussed earlier in \multirefabrev~\cite{canonicalddftfromdecomposition,canonicalddft2,orderpreservingwittmann}.

\bibliography{lit}

\end{document}